\documentclass[aps,prd,onecolumn,groupedaddress,showpacs,nofootinbib,amssymb]{revtex4}
\usepackage[dvips]{graphicx}
\usepackage{amssymb}
\usepackage{amsmath}
\usepackage{graphicx,,color}
\usepackage{amsfonts}
\usepackage{bm}
\usepackage{cancel}
\usepackage{comment}

\newcommand\be{\begin{equation}}
\newcommand\ee{\end{equation}}

\newcommand\e{\mathrm{e}}

\allowdisplaybreaks[4]

\begin{document}

\title{Reheating in Constant-roll $F(R)$ Gravity}
\author{
V.K. Oikonomou,$^{1,2}$\,\thanks{v.k.oikonomou1979@gmail.com}}
\affiliation{ $^{1)}$ Laboratory for Theoretical Cosmology, Tomsk
State University of Control Systems
and Radioelectronics (TUSUR), 634050 Tomsk, Russia\\
$^{2)}$ Tomsk State Pedagogical University, 634061 Tomsk, Russia\\
}

\begin{abstract}
In this work we address the reheating issue in the context of $F(R)$
gravity, for theories that the inflationary era does not obey the
slow-roll condition but the constant-roll condition is assumed. As
it is known, the reheating era takes place after the end of the
inflationary era, so we investigate the implications of a
constant-roll inflation era on the reheating era. We quantify our
considerations by calculating the reheating temperature for the
constant-roll $R^2$ model and we compare to the standard reheating
temperature in the context of $F(R)$ gravity. As we demonstrate, the
new reheating temperature may differ from the standard one, and in
addition we show how the reheating era may restrict the
constant-roll era by constraining the constant-roll parameter.
\end{abstract}

\pacs{04.50.Kd, 95.36.+x, 98.80.-k, 98.80.Cq,11.25.-w}

\maketitle

\section{Introduction}

One of the challenges in modern cosmology is to describe the era
which connects the end of the inflationary period with the
subsequent radiation and matter domination eras, an era known as
reheating era. The Universe after the inflationary era is very cold,
due to the abrupt nearly de Sitter expansion, and the matter
contained in the Universe needs somehow to be thermalized. It is
conceivable that such a task is quite complicated and there exist
various proposals in the literature, see for example the recent
review \cite{Amin:2014eta}. The reheating era invokes many
procedures that are involved in what is now known as Big Bang
Nucleosynthesis, and during this era the energy density that drove
the quasi de Sitter expansion during the inflationary era, will
thermalize the matter content of the Universe. In most well-known
approaches in reheating, the inflaton field transfers its energy to
the Standard Model particles via direct couplings to these fields
\cite{Kofman:1997yn,Greene:1997fu,Kofman:1994rk}. However, in this
context there are some drawbacks of having to fine-tune the
couplings significantly, in order to avoid large couplings during
the inflationary era. An alternative approach for describing the
reheating era is offered by modified gravity, and particularly from
$F(R)$ gravity \cite{Mijic:1986iv}, in which case the reheating
effects take place once the inflationary era ends. In the modified
gravity description, gravity has an effect of the effective equation
of state of the matter fields, and in effect, these produce a
non-trivial effect on the field equations, which in turn thermalize
the matter content of the Universe, see Ref. \cite{Mijic:1986iv} for
the $R^2$ model description of the reheating era.

In most approaches on the description of the reheating era, a
slow-roll era is assumed for the preceding inflationary era.
Recently however, another interesting research stream described an
alternative evolutionary possibility for the inflationary period,
know as constant-roll inflation
\cite{Inoue:2001zt,Tsamis:2003px,Kinney:2005vj,Tzirakis:2007bf,
Namjoo:2012aa,Martin:2012pe,Motohashi:2014ppa,Cai:2016ngx,
Motohashi:2017aob,Hirano:2016gmv,Anguelova:2015dgt,Cook:2015hma,
Kumar:2015mfa,Odintsov:2017yud,Odintsov:2017qpp,Gao:2017owg}, see
also \cite{Lin:2015fqa,Gao:2017uja} for an alternative perspective
on this issue, and also see Ref. \cite{Nojiri:2017qvx} for the
$F(R)$ gravity generalization. The constant-roll inflationary models
have the interesting property of predicting non-Gaussianities
\cite{Chen:2010xka}, even in the context of the single scalar field
models
\cite{Inoue:2001zt,Tsamis:2003px,Kinney:2005vj,Tzirakis:2007bf,
Namjoo:2012aa,Martin:2012pe,Motohashi:2014ppa,Cai:2016ngx,
Motohashi:2017aob}, so this makes these models conceptually
appealing and robust towards future observations of
non-Gaussianities in the power spectrum of primordial curvature
perturbations.

In a recent work we provided a generalization of the constant-roll
inflationary era in the context of $F(R)$ gravity
\cite{Nojiri:2017qvx}, see also \cite{Motohashi:2017vdc} for an
alternative approach. The focus in this letter is to investigate
what are the effects of a constant-roll inflationary era on the
reheating process, in the context of $F(R)$ gravity. We shall use
the $R^2$ model \cite{Starobinsky:1980te,Barrow:1988xh} and we shall
calculate the reheating temperature for the case that the preceding
inflationary era was a constant-roll one, and we will compare the
resulting reheating temperature to the one corresponding to the case
that the inflationary era was a usual slow-roll one. As we
demonstrate, the ratio of the two reheating temperatures can be
quite large, depending strongly on the $F(R)$ gravity model, and
also we show that the reheating era can constraint the constant-roll
era, since the parameters that quantify the constant-roll era must
be constrained for consistency.

This paper is organized as follows: in section II we provide some
essential information for the $F(R)$ constant-roll inflationary era.
We focus on the $R^2$ inflation case, and as we show, the
constant-roll condition affects the rate of the quasi-de Sitter
expansion. In section III we investigate the effects of the
constant-roll quasi-de Sitter expansion on the reheating era, and we
calculate the ratio of the reheating temperatures corresponding to
the constant-roll and slow-roll preceding inflationary eras. As we
show, the ratio depends on the constant-roll parameter that
quantifies the constant-roll era. Finally, the conclusions follow at
the end of the article.

\section{The Constant-roll Inflation Condition with $R^2$
Gravity}

The constant-roll $F(R)$ gravity inflationary era was introduced in
\cite{Nojiri:2017qvx} (see also \cite{Motohashi:2017vdc} for an
alternative viewpoint), and the main assumption was that the
constant-roll condition becomes as follows,
\begin{equation}
\label{constantrollcondition} \frac{\ddot{H}}{2H\dot{H}}\simeq
\beta\, ,
\end{equation}
with $\beta$ being a real parameter. The condition
(\ref{constantrollcondition}) is a natural generalization of the
constant-roll condition for a canonical scalar field, since the
expression in (\ref{constantrollcondition}) is the second slow-roll
index in the $F(R)$ gravity frame. For the purposes of this paper,
we shall consider a vacuum $F(R)$ gravity (for reviews see
\cite{reviews1,reviews2,reviews3}), with the gravitational action
being of the following form,
\begin{equation}
\label{JGRG7} S_{F(R)}= \int d^4 x \sqrt{-g} \left(
\frac{F(R)}{2\kappa^2} \right)\, ,
\end{equation}
with $g$ being the trace of the background metric, which we shall
assume to be a flat Friedmann-Robertson-Walker (FRW) metric with
line element,
\begin{equation}
\label{metricfrw} ds^2 = - dt^2 + a(t)^2 \sum_{i=1,2,3}
\left(dx^i\right)^2\, ,
\end{equation}
where $a(t)$ is the scale factor. Upon variation of the action
(\ref{JGRG7}) with respect to the metric tensor, the gravitational
equations of motion are,
\begin{align}
\label{eqnmotion1}
3F_RH^2=& \frac{F_RR-F}{2}-3H\dot{F}_R \, , \\
\label{eqnmotion2} -2F_R\dot{H}=& \ddot{F}-H\dot{F} \, ,
\end{align}
where the expression $F_R$ stands for $F_R=\frac{\partial
F}{\partial R}$ and the ``dot'' indicates differentiation of the
corresponding quantity with respect to the cosmic time. In Ref.
\cite{Nojiri:2017qvx}, we explored the inflationary dynamics of
constant-roll inflation in the context of $F(R)$ gravity, and as we
showed it is possible to obtain observational indices of inflation
compatible with the Planck observational data. We used two
well-known $F(R)$ gravity models in order to exemplify our results,
and particularly the $R^2$ model and a power-law $F(R)$ gravity
model, and in this paper we shall use the $R^2$ inflation model in
order to investigate how the reheating era is modified if a
constant-roll inflationary era precedes the reheating era. The
$F(R)$ gravity function in the case of the $R^2$ model
\cite{Starobinsky:1980te} has the following form,
\begin{equation}
\label{r2inflation} F(R)=R+\frac{1}{36H_i}R^2\, ,
\end{equation}
with $H_i$ being a phenomenological parameter with dimensions of
mass$^2$, and we assume that $H_i\gg 1$. During the inflationary era
we shall assume that the first slow-roll index
$\epsilon_1=-\frac{\dot{H}}{H^2}$ satisfies $\epsilon_1\ll 1$, and
also that the constant-roll condition (\ref{constantrollcondition})
holds true. Hence, during the era for which the conditions
$\dot{H}\ll H^2$ and also $\ddot{H}\sim 2\beta H\dot{H}$ hold true,
the gravitational equations (\ref{eqnmotion1}) and
(\ref{eqnmotion2}) can be written as follows,
\begin{equation}
\label{frweqnsr2} \ddot{H}-\frac{\dot{H}^2}{2H}+3H_iH=-3H\dot{H}\,
,\quad \ddot{R}+3HR+6H_iR=0\, .
\end{equation}
In view of the constant-roll condition $\ddot{H}\sim 2\beta
H\dot{H}$, the first differential equation appearing in Eq.
(\ref{frweqnsr2}), can be written as follows
\begin{equation}
\label{rsquarebasic} \dot{H}H \left( 2\beta+\frac{\epsilon_1}{2}+3
\right)\dot{H}=-3H_i\, ,
\end{equation}
and due to the fact that $\epsilon_1\ll 1$, by eliminating the
$\epsilon_1$-dependent term in Eq. (\ref{rsquarebasic}), and by
solving the resulting differential equation, we obtain the following
solution at leading order,
\begin{equation}
\label{hubblersquare} H(t)=H_0-H_I(t-t_k)\, ,
\end{equation}
with the parameter $H_0$ being arithmetically of the order
$\mathcal{O}(H_i)$. Also, the parameter $H_I$ appearing in Eq.
(\ref{hubblersquare}) is equal to,
\begin{equation}
\label{hI} H_I=\frac{3H_i}{2\beta+3}\, ,
\end{equation}
and in addition, the time instance $t=t_k$ corresponds to the
horizon crossing time instance. Clearly, the cosmological evolution
(\ref{hubblersquare}) is a quasi-de Sitter evolution, just as in the
ordinary slow-roll $R^2$ inflation model, with the difference being
that in the ordinary $R^2$ model case, $\beta=0$ and hence $H_I\to
H_i$. As it was shown in \cite{Nojiri:2017qvx}, the constant-roll
inflationary era for the $R^2$ model comes to an end, due to the
production of curvature fluctuation, which ends the inflationary
era. Then, when the constant-roll era ends, the constant-roll
condition does not hold true and due to the presence of the term
$\ddot{H}$ in the gravitational equations, the oscillating reheating
era commences, as we evince in the next section.

\section{Reheating in Constant-roll $R^2$ Gravity}

After the graceful exit from the inflationary era, the Universe
enters an intermediate era, which should make a connection between
the inflationary era and the radiation and matter domination eras.
During the reheating era, the Standard Model particles are
thermalized by the Universe, and this is an important feature of any
viable cosmological, since after the inflationary era the Universe
is cold due to the abrupt nearly exponential expansion that the
Universe underwent during the inflationary era. As we mentioned, we
shall investigate how the reheating era is affected due to a
constant-roll $F(R)$ gravity era, and we directly compare the
resulting picture with the ordinary slow-roll $F(R)$ gravity model.
We shall focus on the $R^2$ model of Eq. (\ref{r2inflation}),
although it is expected that similar results can be obtained for any
$F(R)$ gravity. In the following we adopt the approach and notation
of Ref. \cite{Mijic:1986iv}. As we show, the effects of a
constant-roll quasi-de Sitter evolution can be found directly on the
reheating temperature, so we shall make a comparison of the ordinary
slow-roll case and the constant-roll case. The reheating era brings
new cosmological features into play since the term $\ddot{R}$ in the
corresponding differential equation in Eq.~(\ref{frweqnsr2}), cannot
be omitted. In this case, the scalar curvature evolves as a damped
oscillation, with a restoring force being of the form $\sim 3H_i$.
During the reheating era, the Hubble rate can be found by solving
the following differential equation,
\begin{equation}
\label{maindiffhubble}
\ddot{H}-\frac{\dot{H}^2}{2H}+3H_iH=-3H\dot{H}\, .
\end{equation}
During the reheating era the terms $\sim \ddot{H}$ and $\sim
\frac{\dot{H}^2}{2H}$ start to dominate the evolution, however the
term $\sim \dot{H}H$ is comparably negligible. Consider that $t=t_r$
is the time instance that the reheating era commences, so for
$t<t_r$, the Hubble rate is given in Eq. (\ref{hubblersquare}), and
for $t>t_r$, by solving the differential equation
(\ref{maindiffhubble}), we get the following solution,
\begin{equation}
\label{htsolutionreheating} H(t)\simeq \frac{\cos^2\omega
(t-t_r)}{\frac{3}{\omega}+\frac{3}{4}(t-t_r)+\frac{3}{8\omega}\sin
2\omega (t-t_r)}\, .
\end{equation}
The corresponding scale factor during the reheating is,
\begin{equation}
\label{reheatingscalefactor} a(t)=a_r\left(1+\frac{\omega
(t-t_r)}{4} \right)^{2/3}\, ,
\end{equation}
with $a_r$ being equal to $a_r=a_0
\e^{\frac{H_0^2}{2H_I}-\frac{1}{12}}$, and $a_0$ is the scale factor
at the beginning of inflation. In Eq. (\ref{htsolutionreheating}),
the parameter $\omega$ is affected by the transition from
constant-roll to reheating at $t=t_r$, and can be determined by
using the following condition,
\begin{equation}
\label{condition1equal} \left|\frac{\dot{H}^2}{2H}\right| = \left|
3H\dot{H} \right|\, .
\end{equation}
The condition (\ref{condition1equal}) combined with the quasi-de
Sitter evolution (\ref{hubblersquare}) and with the reheating
evolution (\ref{htsolutionreheating}), yields the following form of
$\omega$ and $t_r$,
\begin{equation}
\label{omegareheating} \omega=\sqrt{\frac{3H_I}{2}}\, \quad
t_r\simeq H_I H_0\, .
\end{equation}
An approximate form of the scalar curvature can also be determined,
since during the reheating era, the Ricci scalar is approximately
equal to $R\simeq 6\dot{H}$, thus we approximately have,
\begin{equation}
\label{approxscalarcurvaturereheating} R(t)\simeq -\frac{6\omega
\sin 2\omega
(t-t_r)}{\left(\frac{3}{\omega}+\frac{3}{4}(t-t_r)+\frac{3}{8\omega}\sin
2\omega (t-t_r) \right)}\, .
\end{equation}
At this point recall that the parameter $H_I$ contains a hidden
$\beta$-dependence, as it can be seen in Eq. (\ref{hI}), and recall
that the parameter $\beta$ is the constant-roll condition parameter
of Eq. (\ref{constantrollcondition}). Essentially the parameter
$\beta$ determines the shape and the size of the reheating phase.

Now let us quantify the effects of the constant-roll on the $R^2$
gravity reheating by comparing the reheating temperature for the
constant-roll and slow-roll $R^2$ model. In order to calculate the
reheating temperature, it is assumed that the matter content
consists of a scalar field $\phi$ with gravitational equation
$g^{\mu \nu}\phi_{;\mu \nu}=0$. As it was shown in
\cite{Mijic:1986iv}, the effect of the matter content is connected
to the square average of the scalar curvature $R$, and the $(t,t)$
component of the field equations yields,
\begin{equation}
\label{eqnmotionreheatingmain}
H^2+\frac{1}{18H_I}\left(HR-\frac{R^2}{12}RH^2
\right)=\frac{8\pi}{3}G\rho_c\, ,
\end{equation}
where energy-density term $\rho_c$ is defined to be,
\begin{equation}
\label{rhomatter}
\rho_c=\frac{N}{a^4}\int_{t_r}^t\frac{\omega}{1152\pi} \bar{R}^2a^4
d t\, ,
\end{equation}
and it refers to the constant-roll case. By assuming that during the
reheating era, the energy density $\rho_c$ is totally converted to
radiation energy $\rho_r=\frac{N\pi^2}{30}T_{rc}^4$, where $N$ is
the total number of relativistic particles, while $T_{rc}$ is the
reheating temperature corresponding to the case that a constant-roll
case precedes the reheating epoch. By denoting $\rho_s$ and $T_{rs}$
the energy density and the reheating temperature corresponding to
the case that a slow-roll inflationary era precedes the reheating
epoch, by using Eq. (\ref{rhomatter}) and the corresponding formula
for the slow-roll pre-reheating era, and also by assuming that the
final time instance in the integral appearing in Eq.
(\ref{rhomatter}), is  $t\simeq t_r+10\omega $, we obtain the
following relations,
\begin{equation}
\label{reheatingtemperature1}
\frac{\rho_c}{\rho_s}=\frac{\omega^4}{\omega_s^4},\,\,\,
\frac{T_{rc}}{T_{rs}}=\frac{\omega}{\omega_s}\, ,
\end{equation}
where the parameter $\omega_s$ for the slow-roll era quasi-de Sitter
evolution reads,
\begin{equation}\label{omegas}
\omega_s=\sqrt{\frac{3H_i}{2}}\, .
\end{equation}
By combining Eqs. (\ref{hI}), (\ref{omegareheating}) and
(\ref{omegas}), we finally obtain the ratio of the constant-roll to
slow-roll reheating temperatures,
\begin{equation}\label{finalrelation}
 \frac{T_{rc}}{T_{rs}}=\sqrt{\frac{3}{2 \beta +3}}\, ,
\end{equation}
in which it can clearly be seen that the parameter $\beta$ affects
the reheating temperature, so the constant-roll era preceding the
reheating era, can affect the reheating era.
\begin{figure}[h]
\centering
\includegraphics[width=18pc]{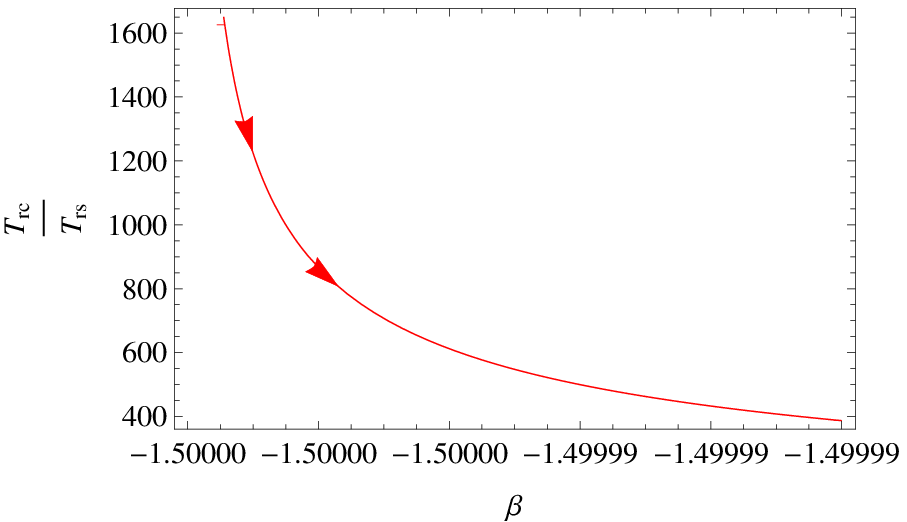}
\includegraphics[width=18pc]{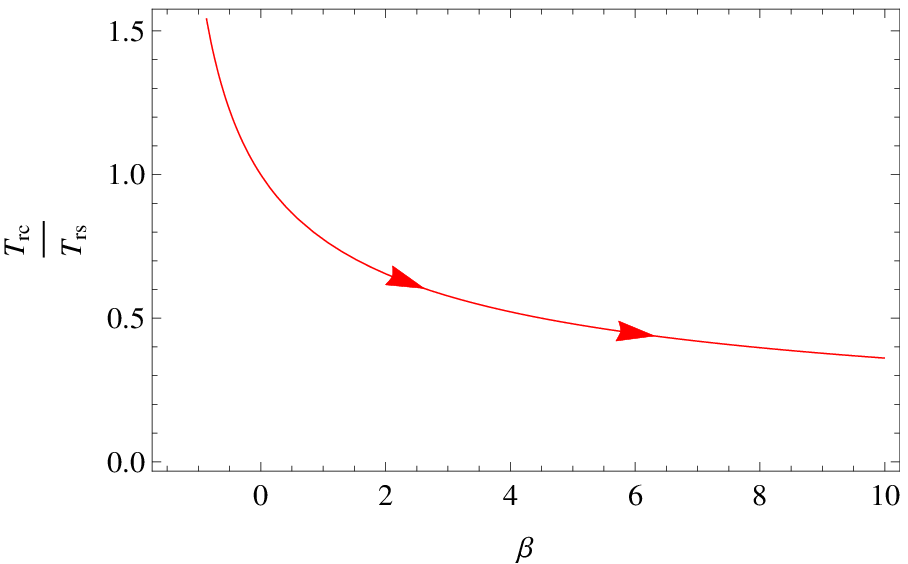}
\caption{Behavior of the ratio $\frac{T_{rc}}{T_{rs}}$ as a function
of the constant-roll parameter $\beta$.}\label{plot1}
\end{figure}
In order to have a quantitative picture of how the constant-roll
parameter $\beta$ affects the reheating temperature, we shall plot
the ratio $\frac{T_{rc}}{T_{rs}}$, as a function of $\beta$. In Fig.
\ref{plot1}, we plot the behavior of the ratio
$\frac{T_{rc}}{T_{rs}}$ as a function of the constant-roll parameter
$\beta$. As it can be seen in the left plot, the ratio
$\frac{T_{rc}}{T_{rs}}$ takes large values as $\beta\to -3/2$, since
the ratio is singular at $\beta=-3/2$. This means that for $\beta\to
-3/2$, the constant-roll reheating temperature may differ
significantly from the corresponding slow-roll reheating
temperature. However, as $\beta$ takes larger values, the ratio
$\frac{T_{rc}}{T_{rs}}$ tends to zero, which means that the
reheating temperature in the two cases is almost the same. Moreover,
the equation (\ref{finalrelation}) can be considered as a constraint
on the parameter $\beta$ for the constant-roll $R^2$ model, since
$\beta$ must be $\beta>-3/2$, so the reheating era imposes some
constraints on the constant-roll inflationary era.

\section{Conclusions}

In this paper we discussed the implications of a constant-roll
inflationary era on the dynamics of reheating, in the context of
$F(R)$ gravity. After discussing how the constant-roll condition
modifies the dynamics of inflation in $F(R)$ gravity, we focused on
the $R^2$ model and we demonstrated that the constant-roll condition
predicts a quasi-de Sitter evolution which governs the inflationary
era. As we demonstrated this quasi-de Sitter evolution affects the
reheating era, and it modifies directly the reheating temperature.
We compared the reheating temperature corresponding to a
constant-roll inflationary era preceding the reheating era, to the
corresponding slow-roll case for the $R^2$ model, and we found that
when the constant-roll parameter approaches $\beta\to -3/2$, the two
reheating temperatures may differ significantly. Finally, we
demonstrated that the reheating era may restrict the constant-roll
parameter and in effect the whole constant-roll inflationary era,
but this feature is strictly model dependent. For example, in the
case of the $R^2$ model, the constant-roll parameter must be
$\beta>-3/2$ for physical consistency reasons. In principle the
results we found in this work may apply to other $F(R)$ gravities as
well, and it would be interesting to investigate this issue in the
context of string inspired modified gravities, such Gauss-Bonnet
scalar theories, a task we hope to address in a future work.

\section*{Acknowledgments}

This work is supported by the Russian Ministry of Education and
Science, Project No. 3.1386.2017 (V.K.O).

\end{document}